\documentclass[twocolumn]{aa}
\usepackage{url}
\usepackage{graphicx}
\usepackage{txfonts}
\usepackage{color}
\usepackage{hyperref}

\definecolor{Mygray}{gray}{0.4}

\begin{document}

\title{Comptonization of the cosmic microwave background by high energy particles residing in AGN cocoons}

\author{D. A. Prokhorov\inst{1}, V. Antonuccio-Delogu\inst{2,3}, Joseph Silk\inst{4,5}}

\offprints{D. A. Prokhorov \email{phdmitry@gmail.com}}

\institute{Korea Astronomy and Space Science Institute, Hwaam-dong,
Yuseong-gu, Daejeon, 305-348, Republic of Korea \and
            INAF-Osservatorio Astrofisico di Catania, Via S. Sofia 78, I-95123
Catania, Italy \and Scuola Superiore di Catania, Via San Nullo 5/i,
Catania, I-95123, Italy \and Astrophysics, Department of Physics,
University of Oxford, Keble Road Ox1 3RH, Oxford, United Kingdom \and
Institut d'Astrophysique de Paris, 98bis Blvd Arago, Paris 75014, France }

\authorrunning{Prokhorov et al.}
\titlerunning{Comptonization of the CMB by particles residing in AGN cocoons}

\abstract
{X-ray cavities and extended radio sources (`cocoons') surrounding
active galactic nuclei (AGN) have been detected by the Chandra X-ray
mission and radio interferometers. A joint analysis of X-ray and
radio maps suggests that pressure values of non-thermal
radio-emitting particles derived from the radio maps are not
sufficient to inflate the X-ray cavities. We propose using the
Sunyaev-Zel'dovich (SZ) effect, whose intensity strongly depends on
the pressure, to find the hitherto undetected, dynamically-dominant
component in the radio cocoons.}
{Numerical simulations show that plasma with a high temperature
($10^9 - 10^{10}$ K) is a good candidate for inflating the AGN cocoons. To constrain the
population of high energy electrons inside AGN cocoons that is predicted by
numerical simulations, we study different methods for maximizing the
contribution of such energetic electrons to the SZ effect.}
{Our calculations of intensity maps of the SZ effect include
relativistic corrections and utilize both analytic models and
numerical 2-D simulations.}
{We demonstrate that the spectral function at a frequency of 217 GHz
has an absolute maximum at a temperature higher than $10^9$ K,
therefore the measurement of the SZ effect at this frequency is a
powerful tool for potentially revealing the dynamically-dominant
component inside AGN jet-driven radio cocoons. A new method is
proposed for excluding the contribution from the low energy,
non-relativistic electrons to the SZ effect by means of observations
at two frequencies.  We show how one may correct for a possible
contribution from the kinematic SZ effect. The intensity maps of the
SZ effect are calculated for the self-similar Sedov solution, and
application of a predicted ring-like structure on the SZ map at a
frequency of 217 GHz is proposed to determine the energy released
during the active jet stage. The SZ intensity map for an AGN cocoon
in a distant elliptical is calculated using a 2-D numerical
simulation and including relativistic corrections to the SZ effect.
We show the intensity spectrum of the SZ effect is flat at high
frequencies if gas temperature is as high as $k_\mathrm{b}
T_{\mathrm{e}}=500$ keV.}
{}


\maketitle

\maketitle

\section{Introduction}

Inverse Compton scattering of the cosmic microwave background
radiation (CMB) by free electrons in intervening ionized gas allows
us to study the physical state of the ionized gas (for a review, see
Sunyaev \& Zel'dovich 1980). The Comptonization process of a soft
photon spectrum by a Maxwellian distribution of non-relativistic
electrons is described by the Kompaneets equation (Kompaneets 1957).
An analytical solution of the Kompaneets equation was derived by
Zel'dovich \& Sunyaev (1969) and applied to the distortion of the
cosmic microwave background radiation by Thompson scattering in
clusters of galaxies (the Sunyaev-Zel'dovich effect, hereinafter the
SZ effect). The SZ effect has proven to be an important tool for
cosmology and the study of clusters of galaxies (for a review, see
Birkinshaw 1999). It measures the pressure of an electron population
integrated along the line of sight and does not depend on spectral
features of the underlying electron distributions, as long as they
are non-relativistic.

A relativistically correct formalism based on the probability
distribution of the photon frequency shift after scattering was
given by Wright (1979) to describe the Comptonization process of
soft photons by mildly relativistic electrons. The Wright formalism
(as is the Kompaneets approximation) is based on two assumptions:
(1) The Thomson cross-section is applicable; and (2) the photons are
sufficiently soft that the energy of the scattered photons is less
than that of the electrons. The validity regime of the Wright
formalism was studied by Loeb et al. (1991) who have shown that this
formalism is valid for many astrophysical environments. The
relativistically correct Wright formalism was used to calculate the
SZ effect in galaxy clusters by many authors (for a review, see
Rephaeli 1995 and Birkinshaw 1999).

Although the SZ effect is observed in clusters of galaxies, there
are other environments in which this effect should be observable by
the next generation of telescopes with high sensitivity and good
angular resolution (such as the ALMA telescope
\footnote{http://www.eso.org/sci/facilities/alma/science/cosmo.html}).
Since a significant fraction of the energy released by active
galactic nuclei (AGN) goes into heating of the neighbouring plasma
by shocks,  distortion of the CMB due to the SZ effect by
shock-heated plasma may be produced.

According to the standard evolutionary scenario for strong double
radio sources (Scheuer 1974; Blandford \& Rees 1974), jets boring
through the intergalactic medium (IGM) are not in direct contact
with the undisturbed IGM, but rather are enveloped in a cocoon
consisting of shocked jet material and shocked IGM. Scheuer (1974)
showed that the cocoon around a pair of supersonic, low-density
(when compared to the ambient IGM) jets acts as a ``wastebasket''
for most of the energy deposited by the jets. Begelman \& Cioffi
(1989) argued that the cocoons in many observed sources have not yet
had time to reach pressure balance with the ambient medium.

The Comptonization of the CMB by the hot plasma around strong radio
sources was considered in the framework of the non-relativistic
Kompaneets approximation by Yamada et al. (1999), Platania et al.
(2002) and Cavaliere \& Lapi (2006). However, numerical simulations
by Antonuccio-Delogu \& Silk (2008) and Tortora et al. (2009) show
that the temperatures can become very high within the cocoons (T
$\approx 10^9-10^{11}$ K), and therefore consideration of the
relativistically correct SZ formalism is necessary. Kino et al.
(2007) found that thermal electron temperatures of the AGN cocoons
are predicted to be of the order of an MeV and determined only by
the bulk Lorentz factor of the jet. Therefore, the relativistic
Wright formalism must be applied to  correctly derive the distortion
of the CMB by AGN cocoons. The temperatures in AGN cocoons are
orders of magnitude higher than those of clusters of galaxies
($T_{\mathrm{cl}}\approx 10^7-10^8$ K).

Recently, numerous X-ray cavities in the intra-cluster medium have
been detected by the Chandra X-ray observatory (e.g. McNamara et al.
2000; McNamara et al. 2005). They often coincide with the radio
lobes of the central radio galaxy. The non-thermal pressure derived
from the equipartition condition for the energy of
synchrotron-radiating non-thermal electrons and magnetic fields is a
factor of ten smaller than the pressures required to inflate the
bubbles (e.g. Ito et al. 2008). This implies that most of the
energy in the cocoon is carried by an invisible component such as
high energy thermal electrons (e.g. Ito et al. 2008). Observations
of the SZ effect were proposed by Pfrommer et al. (2005) to probe
the inferred dynamically-dominant component of plasma bubbles
associated with X-ray cavities. Pfrommer et al. (2005) studied
examples of several different physical scenarios concerning the composition
of the plasma bubble which is as a whole in pressure equilibrium
with the ambient ICM, and derived the SZ effect due to electrons in
the plasma bubbles using the Wright formalism. However, cocoons are
expected to be over-pressured with respect to the ambient IGM
(Begelman \& Cioffi 1989), and therefore the SZ effect derived under
the assumption of  pressure balance gives only a lower limit on the
true SZ effect produced in the cocoons.

To constrain a population of high energy electrons inside AGN
cocoons predicted by numerical simulations, we study in this paper
the induced CMB distortion as a function of gas temperature. Taking into
account the derived temperature dependences, we consider different
methods for maximizing the contribution of such energetic electrons.
By using the self-similar Sedov solution and the results of our
numerical simulations, we calculate the relativistically correct SZ
intensity maps.

The plan of the paper is as follows. We consider the SZ effect by
mildly relativistic electrons in the framework of the Wright
formalism in Sect. 2 to find a convenient method for observing an
electron component with high temperatures ($T\gg10^8$ K) derived
from numerical simulations of AGN cocoons. The Sedov stage of the
AGN cocoon evolution is considered in Sect. 3, where intensity maps
of the SZ effect are calculated. The SZ effect produced by high
energy electrons residing in a simulated AGN cocoon is studied in
Sect. 4. We show in Sect. 5 that the CMB spectral distortion is flat
in the broad high frequency range if gas temperatures are high. We
calculate an intensity map of the SZ effect by means of the
dynamical scaling of cocoon in Appendix A.

\section{Generalized spectral functions.}

In this section, the spectral properties of the SZ effect are studied
within the framework of the Kompaneets and Wright formalisms to
determine the  contribution of the high
temperature electron component.

\subsection{A generalized spectral function}

The CMB spectral distortion produced by the SZ effect in the
Kompaneets approximation is (Zel'dovich \& Sunyaev 1969):

\begin{equation}
\Delta I_{nr}(x) = I_{\mathrm{0}} g(x) y_{\mathrm{gas}} \label{Inr}
\end{equation}
where $x=h\nu/k_{\mathrm{b}} T_{\mathrm{cmb}}$, $I_{\mathrm{0}}=2
(k_{\mathrm{b}} T_{\mathrm{cmb}})^3 / (hc)^2$, and the spectral
function $g(x)$ is given by:
\begin{equation}
g(x)=\frac{x^4 \exp(x)}{(\exp(x)-1)^2} \left(x
\frac{\exp(x)+1}{\exp(x)-1}-4\right).
\end{equation}
The subscript  $`nr' $ denotes the fact that Eq. (\ref{Inr}) was
obtained in the non-relativistic limit.
The Comptonization parameter $y_{\mathrm{gas}}$ is given by
\begin{equation}
y_{\mathrm{gas}}=\frac{\sigma_{\mathrm{T}}}{m_{\mathrm{e}}c^2} \int
dl n_{\mathrm{gas}} k_{b} T_{\mathrm{e}}
\end{equation}
where the line-of-sight integral extends from the last scattering
surface of the CMB radiation to the observer at redshift z=0,
$T_{\mathrm{e}}$ is the electron temperature, $n_{\mathrm{gas}}$ is
the number density of the gas, and $\sigma_{\mathrm{T}}$ is the
Thomson cross-section, $m_{\mathrm{e}}$ the electron mass, $c$ the
speed of light, $k_{\mathrm{b}}$ the Boltzmann constant and $h$ the
Planck constant.

The CMB spectral distortion in the relativistically corrected formalism
can be written in an analogous form to that given by
(En$\ss$lin \& Kaiser 2000; Pfrommer et al. 2005), and is

\begin{equation}
\Delta I(x) = I_{\mathrm{0}}
\frac{\sigma_{\mathrm{T}}}{m_{\mathrm{e}}c^2} \int dl
n_{\mathrm{gas}} k_{\mathrm{b}} T_{\mathrm{e}} G(x, T_{\mathrm{e}}).
\label{form}
\end{equation}
Note that in our notation, the spectral function $g(x)$ is changed to the
generalized spectral function $G(x, T_{\mathrm{e}})$ which depends
explicitly on the electron temperature.

The relativistic spectral function $G(x, T_{\mathrm{e}})$ derived in
the framework of the Wright formalism is given by
\begin{equation}
G(x, T_{\mathrm{e}})=\int^{\infty}_{-\infty} \frac{P_{1}(s,
T_{\mathrm{e}})}{\Theta(T_{\mathrm{e}})} \left(\frac{x^3 \exp(-3
s)}{\exp(x \exp(-s))-1}-\frac{x^3}{\exp(x)-1}\right) ds \label{G}
\end{equation}
where $\Theta(T_{\mathrm{e}}) = k_{\mathrm{b}} T_{\mathrm{e}}/
m_{\mathrm{e}}c^2$, $P_{1}(s, T_{\mathrm{e}})$ is the distribution
of frequency shifts for single scattering (Wright 1979; Birkinshaw
1999).

The following limiting cases hold:
\begin{enumerate}
\item
if $T_{\mathrm{e}}\ll 10^8$ K, the values of relativistic $G(x,
T_{\mathrm{e}})$ and non-relativistic $g(x)$ spectral functions are
equal;
\item
if $T_{\mathrm{e}}\gg 10^{10} $K, the value of the relativistic
spectral function is written as
\begin{equation}
G(x, T_{\mathrm{e}})\approx-\frac{x^3}{\exp(x)-1}
\frac{m_{\mathrm{e}}c^2}{k_{\mathrm{b}} T_{\mathrm{e}}} \label{rel}
\end{equation}
where $i(x)=-x^3/(\exp(x)-1)$ is the Planck spectrum.
\end{enumerate}
Note that in the papers by En$\ss$lin \& Kaiser (2000) and Pfrommer
et al. (2005), a different form for the  relativistically correct
Wright formalism is proposed. These authors have chosen to average
over the gas temperature along the line of sight. The two forms
coincide if the astrophysical medium is homogeneous. However, if the
astrophysical medium is inhomogeneous, use of the form given by Eq.
(\ref{form}) is more convenient since this uses local temperature
values.

In order to study the properties of the generalized spectral function $G(x,
T_{\mathrm{e}})$, we calculate this function at frequencies 90 GHz
and 217 GHz at which the SZ effect will be observed by the ALMA
telescope. These frequencies are also observable by other telescopes
(like the South Pole Telescope and the Atacama Cosmology Telescope)
and are motivated by the following reasons: 90 GHz is close to the
frequency of the SiO (silicon monoxide) line, and 217 GHz
corresponds to the crossover frequency of the non-relativistic SZ
effect.

Pfrommer et al. (2005) simulated a Green Bank Telescope (GBT)
observation of the central regions of the Perseus and Abell 2052
clusters where X-ray cavities have been detected by the Chandra observatory, and
computed the frequency band-averaged SZ flux decrement in the
frequency interval [86 GHz, 94 GHz]. The ALMA telescope will observe
at a frequency 90 GHz with higher sensitivity and resolution than
that of the GBT; comparison the sensitivity of the GBT with the ALMA
telescope has  already been
studied\footnote{http://safe.nrao.edu/wiki/bin/view/GBT/GBTSensitivityComparison}.

\begin{figure}[h]
\centering
\includegraphics[angle=0, width=7cm]{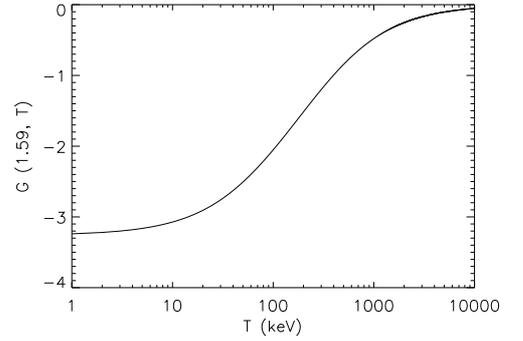}
\caption{Dependence of the generalized spectral function at
 frequency 90 GHz on the plasma temperature.} \label{Fig90}
\end{figure}

In Fig. \ref{Fig90} we show the dependence of the generalized
spectral function $G(x, T_{\mathrm{e}})$ at a frequency 90 GHz
(x=1.59), derived from Eq. (\ref{G}) on temperature. We find that
the absolute value of the generalized spectral function at a
frequency 90 GHz significantly (and monotonically) decreases with
electron temperature. The Kompaneets approximation is valid when the
change of a photon frequency due to Compton scattering is much
smaller than an initial photon frequency $\Delta\nu/\nu\ll1$. Since
the average photon frequency change per inverse Compton scattering
equals $\Delta\nu/\nu=4k_{\mathrm{b}} T_{\mathrm{e}}/(m_{\mathrm{e}}
c^2)$ (for non-relativistic electrons, see, e. g., Rybicki \&
Lightman 1979), the Kompaneets approximation is invalid when
$k_{\mathrm{b}} T_{\mathrm{e}}\simeq (m_{\mathrm{e}}c^2)/4$, i.e.
when $k_{\mathrm{b}} T_{\mathrm{e}}\simeq 128 \mathrm{keV}$. Thus,
strong deviations from the intensity value (and from the spectral
function value of $g(1.59)\approx-3.28$) derived in the Kompaneets
approximation arise at high temperatures. The given estimate of the
regime of breakdown of the Kompaneets approximation is an
approximation and significant deviations (although not of order
unity) take place at much lower temperatures, even at 30 keV (e.g.
Fabbri 1981).

\begin{figure}[h]
\centering
\includegraphics[angle=0, width=7cm]{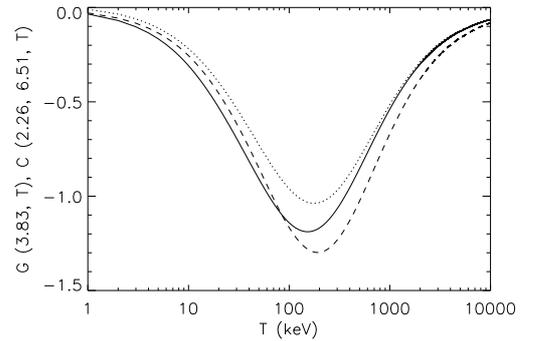}
\caption{Dependences of the generalized spectral function at the
frequency 217 GHz and the combined generalized spectral function
C(2.26, 6.51, T) on  plasma temperature are shown by solid and
dashed lines, respectively.} \label{Fig217}
\end{figure}

An interesting feature of the SZ effect is that at frequency 217 GHz
(x=3.83) where the SZ effect in the framework of the Kompaneets
approximation is zero, the SZ effect from an AGN cocoon is dominated
by the high temperature electron component ($T_{\mathrm{e}}\gg 10^8$
K). Therefore, the SZ effect at frequency 217 GHz is an interesting
tool for analyzing the hot electron component in AGN cocoons. Note
that Colafrancesco (2005) has considered a measurement of the SZ
effect at a frequency 217 GHz to analyze a non-thermal electron
component in radio lobes. Since the value of $G(3.83,
T_{\mathrm{e}})$ is small when the temperature is either much lower
than $10^{8}$ K or higher than $10^{10}$ K, we conclude that this
function should have an absolute maximum at an intermediate
temperature.

Dependence of the generalized spectral function $G(x,
T_{\mathrm{e}})$ at a frequency 217 GHz (x=3.83) on temperature
derived from Eq. (\ref{G}) is shown in Fig. \ref{Fig217}. The
extremum of a curve $G(3.83, T_{\mathrm{e}})$ is at a temperature
$T_{\mathrm{e}}\approx 160$ keV ($\approx 1.9\times10^9$ K) which is
in the temperature range found from numerical simulations of
cocoons. Therefore analysis of the SZ effect at this frequency
should be a promising way of  demonstrating the existence of an
electron component at high temperature. Note that combined
generalized functions defined in Sect. 2.2 are also shown in Fig.
\ref{Fig217}.

Measurements at the frequency of 217 GHz are sensitive to the
spectral response of the detectors. In Appendix B we show how the
broad detector spectral response impacts on the possibility of an
analysis of high temperature plasmas.

In  Sect. 2.2, we investigate the possibility of constraining the properties of high energy particle populations by means of  SZ intensity
measurements at two frequencies.

\subsection{Combined generalized spectral functions}
Measurements of the SZ effect at a frequency 217 GHz is a unique way
of revealing a population of mildly relativistic electrons in AGN
cocoons if the intensity distortion is observed at a single
frequency. We propose below a new method for excluding the
contribution from the low energy, non-relativistic electrons to the
SZ effect by means of observations at two frequencies.

By $C (x_{1}, x_{2}, T_{e})$, we denote a combined generalized
spectral function which corresponds to a relativistic contribution
to the generalized function at a frequency $x_{1}$ taking into
account a measurement at a frequency $x_{2}$ and is defined as
\begin{equation}
C (x_{1}, x_{2}, T_{e})=G(x_{1}, T_{e})-G(x_{2}, T_{e})\times
\frac{g(x_{1})}{g(x_{2})}. \label{C}
\end{equation}
At low temperatures $T_{\mathrm{e}}\ll 10^8$ K, the limiting case
(1) holds (see Sect. 2.1) and, therefore the combined generalized
spectral function $C (x_{1}, x_{2}, T_{e})$ equals 0. It means that
this function is insensitive to low-temperature electron components.
(Note that a SZ intensity cavity at a frequency of 90 GHz produced
by mildly relativistic electrons in an AGN cocoon due to
relativistic SZ effect corrections (Pfrommer et al. 2005) can vanish
if the assumption of pressure equilibrium doesn't hold. This is
because the SZ signal at a frequency of 90 GHz produced by a gas
with a low temperature can be reproduced by a mildly relativistic
gas with a high temperature and higher pressure, see Fig.
\ref{Fig90}). Equation (\ref{C}) defines a family of spectral
functions: each choice of values for frequencies $x_1$, $x_2$
produces a different spectral function. The common property of
spectral functions of this family is that they are more sensitive to
mildly relativistic electron populations than to non-relativistic
electron populations. Therefore, this choice for the combined
generalized spectral function is motivated by studying of high
energy electron populations. Let us consider different choices of
frequencies of $x_1$ and $x_2$.

There are three basic spectral features that characterize the
thermal, non-relativistic SZ effect signal: a minimum of its
intensity located at a dimensionless frequency 2.26 ($\nu$=128 GHz),
a crossover frequency $x_0=3.83$ ($\nu$=217 GHz), and a maximum of
its intensity whose frequency location at a dimensionless frequency
6.51 ($\nu=369$ GHz). If the $x_{1}=3.83$ then the values of
generalized $G(x_1)$ and $C(x_1, x_2)$ are coincident. If
$x_2=3.83$, the combined generalized function is undetermined
because the value of $g(x=3.83)$ equals 0. We noticed that values of
the spectral function of $G(x, T_{\mathrm{e}})$ at high temperatures
at frequencies of $x_1 = 2.26$ and $x_2 = 6.51$ cannot be
simultaneously fitted by the spectral function of $c \times g(x)$,
where $c$ is the arbitrary constant. Therefore, the choice of
frequencies $x_1$ = 2.26 and $x_2$ = 6.51 corresponding to minimum
and maximum values of the SZ intensity in the Kompaneets
approximation is suitable to analyze high energy electron
populations. The dependence of the combined generalized function
with $x_1=2.26$ and $x_2=6.51$ on the plasma temperature is shown in
Fig. \ref{Fig217}. The combined function of $C(1.59, 6.51,
T_{\mathrm{e}})$ where the lower frequency corresponds to 90 GHz is
shown in Fig. \ref{Fig217} by dotted line. Figure \ref{Fig217} shows
that the curves which correspond to the generalized spectral
function at  frequency 217 GHz and the combined generalized function
$C(2.26, 6.51, T)$ are very similar and have peaks at temperatures
higher than 100 keV. Therefore, using the combined generalized
function provides us with an alternative and equivalent method for
studying a population of electrons with energies higher than 100
keV. Such peaks at temperatures higher than 100 keV as that in the
Fig. \ref{Fig217} permit us to maximize the SZ effect from very hot
gas which is expected inside AGN cocoons from numerical simulations.

To calculate the intensity map of the SZ effect using combined
generalized spectral functions, we use Eq. (\ref{form}) where the
generalized spectral function $G(x, T_{\mathrm{e}})$ should be
changed to the combined generalized spectral function $C(x_{1},
x_{2}, T_{\mathrm{e}})$. The kinematic Sunyaev-Zel'dovich effect
which is a possible source of bias in the observations of the SZ
effect by energetic electrons will be considered in Sect. 4.1.

In the next section, an analysis of the SZ effect by means of the
generalized spectral functions is considered for a specific
astrophysical important case - the Sedov self-similar solution.

\section{A specific test case: the Sedov solution.}

Self-similar solutions for a strong point explosion in an ambient
medium are used for modelling adiabatic supernova remnants, solar
flares and processes in active galactic nuclei (Ostriker \& McKee
1988). Sedov (1959) gives the analytical self-similar solution for
description of the motion of a shock front and the distribution of
fluid parameters inside the shocked region for a strong point
explosion in a uniform ambient medium. The gas flow pattern is
determined by only two parameters: the ambient gas density
$\rho_{1}$, and the amount of energy $E$ released in the explosion.
The distance of the shock from the origin is given by
\begin{equation}
R=\beta \left(\frac{E t^2}{\rho_{1}}\right)^{1/5} \label{Rsedov}
\end{equation}
where $\beta$ is a numerical constant.

The density $\rho$ as a function of the dimensionless radial
coordinate $r/R$ decreases rapidly into the sphere and almost all
the gas is in a relatively thin layer behind the shock wave (Sedov
1959). As $r/R\rightarrow0$, the pressure $p$ tends to a constant
limit and the temperature accordingly becomes very high.
Since the SZ effect depends on the thermal energy density of the
electron population along a line of sight, a significant SZ effect
is expected from the central region although the density is very
small there.

Yamada et al. (1999) investigated the SZ effect produced by cocoons
of radio galaxies and constructed a model for the evolution of a
cocoon after the jet is turned off. They examined the evolution of a
cocoon after the jet turns off by analogy with the evolution of a
supernova remnant and showed that the cocoon remains hot enough to
be the source of the SZ effect only during the free expansion and
Sedov (adiabatic) phases. The Sedov stage of the evolution of the
cocoon surrounding an AGN was also considered by Platania et al.
(2002) and Chatterjee \& Kosowsky (2007). In these papers,  for
simplicity the density of the gas inside the cocoon was assumed to
be uniform and only the time evolution of the Comptonization
parameter was taken into account. We will consider a more realistic
case where the gas density depends on the position inside a cocoon.

\begin{figure}[h]
\centering
\includegraphics[angle=90, width=8cm]{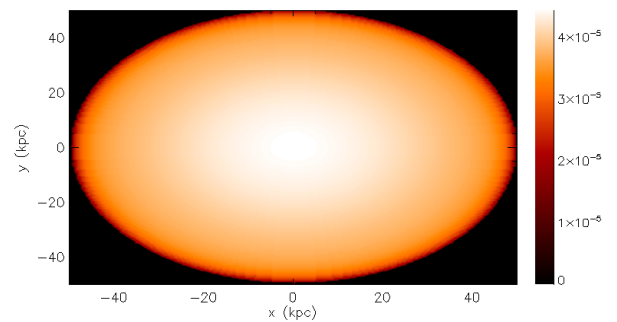}
\caption{The intensity map $|\Delta I/ I_0|$ of the SZ effect
derived from the assumption of the Sedov solution and the Kompaneets
approximation} \label{sz-sed-kom}
\end{figure}

The values of pressure and temperature at any radius inside a cocoon
are determined by the ambient gas density $\rho_{1}$, the amount of
energy $E$ released in the explosion, and time $t$. However, the
parameters directly observed by X-ray telescopes are the ambient
number density $n_{1}=\rho_{1}/m_{\mathrm{p}}$, the shock radius $R$
and the Mach number $M$ of the shock. The Mach number of a shock is
usually derived from the Rankine-Hugoniot jump conditions (for a
review, see Markevitch \& Vikhlinin 2007). Another way to derive the
Mach number of a shock is based on measurements of the flux ratio of
the FeXXV and FeXXVI iron lines (Prokhorov 2010). Since the amount
of energy released and time can be expressed in terms of the shock
radius and the Mach number of the shock, we choose $n_{1}$, $R$, and
$M$ as three parameters for describing the Sedov solution. Typical
values of $n_{1}=10^{-2}$ cm$^{-3}$, $R=50$ kpc, and $M=2.0$
correspond to that of observed from AGN cocoons in galaxy clusters
(Forman et al. 2005; Nulsen et al. 2005a). The Mach number of the
shock can be calculated in terms of the shock velocity relative to
the undisturbed gas given by (Sedov 1959)

\begin{equation}
u=\frac{2 \beta}{5} \left(\frac{E}{\rho_{1} t^3}\right)^{1/5}.
\end{equation}

The self-similar Sedov solution assumes that the pressure $P_2$
behind the shock is much larger than the pressure $P_1$ of the
ambient gas (see Landau \& Lifshitz 1959). The evolution of a
supernova remnant in the high-pressure ambient gas was examined by
Tang \& Wang (2005) by means of high-resolution hydrodynamic
simulations. They found that the supernova remnant evolution
deviates from the Sedov solution when the shock radius is higher
than the characteristic radius $R_{\mathrm{c}}$ (see Eq. 5 in their
paper). Using the typical values of the number density $n_1=0.01$
cm$^{-3}$, the ambient gas temperature $T=2$ keV, and the release
energy $E=3\times10^{60}$ erg, we found that the characteristic
radius for an AGN cocoon equals 80 kpc and higher than the shock
radius $R=50$ kpc adopted above. Therefore, the self-similar Sedov
solution is a reasonable approximation in our case.

Assuming that a short-lived AGN source can be modeled as an
instantaneous central explosion, Cavaliere \& Lapi (2006) calculated
the SZ effect in the framework of the Kompaneets approximation for
the radial pressure profile taken form the Sedov solution. The
intensity map of the SZ effect derived under the assumptions of the
Sedov solution and the Kompaneets approximation is plotted in Fig.
\ref{sz-sed-kom}. We note that in this case, the intensity map of
the SZ effect is almost plane with a slight intensity increase in
the central region.

\begin{figure}[h]
\centering
\includegraphics[angle=90, width=8cm]{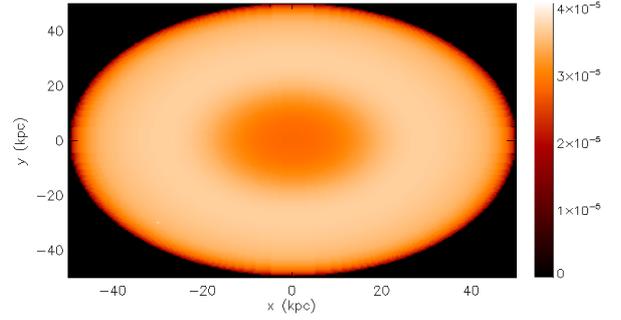}
\caption{The intensity map $|\Delta I/ I_0|$ of the SZ effect at a
frequency 90 GHz derived from the Sedov solution in the framework of
the Wright formalism.} \label{sz90}
\end{figure}

\begin{figure}[h]
\centering
\includegraphics[angle=90, width=8cm]{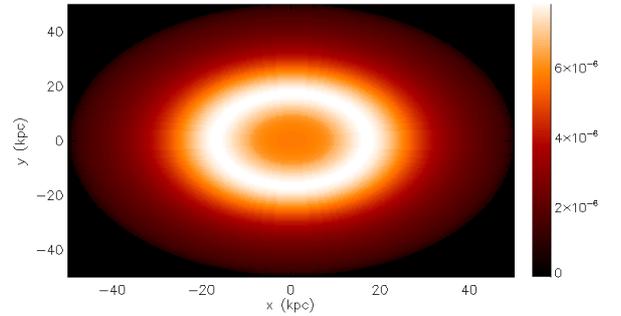}
\caption{The intensity map $|\Delta I/ I_0|$ of the SZ effect at a
frequency 217 GHz derived from the Sedov solution in the framework
of the Wright formalism.} \label{sz217}
\end{figure}

To calculate the SZ intensity maps at frequencies 90 and 217 GHz, we
use the generalized spectral functions derived in the previous
section for different values of temperature (see Eq. \ref{G}, Figs.
1 and 2). Using the Wright formalism and values of the three parameters
introduced above, we calculate the SZ intensity map taking into
account the presence of high temperature gas inside the cocoon. The
intensity maps of the SZ effect at frequencies 90 GHz and 217 GHz
are plotted in Figs. \ref{sz90} and \ref{sz217}, respectively.

Figure \ref{sz90} shows the SZ intensity decrement at a frequency 90
GHz at the center of this map in contrast to the SZ intensity map
derived in the framework of the Kompaneets approximation (see Fig.
\ref{sz-sed-kom}). This central SZ decrement is provided by
electrons with very high temperatures presented in the central
region of the cocoon and the existence of such a decrement is
consistent with the decrease the generalized spectral function at a
frequency 90 GHz with temperature (see Fig. \ref{Fig90}).

The SZ intensity map at a frequency 217 GHz shows a ring-like
structure that is clearly seen in Fig. \ref{sz217}. This ring-like
structure is associated with a SZ signal from a gas layer with
temperature close to $T \approx 160$ keV that corresponds to the
maximum of the absolute value of the general spectral function (see
Fig. \ref{Fig217}). Since the value $G(3.83, 160
\mathrm{keV})\approx -1.2$ of the generalized function at a
frequency 217 GHz and temperature $T \approx 160$ keV is much higher
than that is at much lower and higher temperatures (see Fig.
\ref{Fig217} and Eq. \ref{rel}), the  SZ signal is therefore a
promising tool for observing and analyzing plasma regions with a
temperature close to $2\times10^9$ K.

\subsection{Astrophysical application of the ring at 217 GHz}

In this section,  we consider a way of constraining the amount of energy
$E$ released during the active phase of a jet by using the location of
a ring on the SZ intensity map at a frequency 217 GHz (or on the SZ
intensity map derived from measurements at the 128 and 369 GHz
frequencies, see Sect. 2.2).

The gas temperature at the back of the shock can be expressed in
terms of the shock velocity $u$ as (Sedov 1959)
\begin{equation}
T_2=\frac{2 (\gamma-1) m_{\mathrm{p}}
u^2}{k_{\mathrm{b}}(\gamma+1)^2}
\end{equation}
where $m_{\mathrm{p}}$ is the proton mass and $\gamma$ is the
adiabatic index. The time-dependence of the gas temperature behind
the shock is then
\begin{equation}
T_2=\frac{8 \beta^2 (\gamma-1) m_{\mathrm{p}} E^{2/5}}{25
(\gamma+1)^2 k_{\mathrm{b}} \rho^{2/5}_1 t^{6/5}}.
\end{equation}
The temperature profile inside the sphere (Landau \& Lifshitz 1959)
is written as
\begin{equation}
T_{\mathrm{r}}=T_2 \left(\frac{R}{r}\right)^{3/(\gamma-1)}\label{T2}
\end{equation}
where $r$ is the distance from the center. Although Eq. (\ref{T2})
was derived under the assumption that $r\ll R$, we verify that this
equation can also be applied if $r\lesssim R$. Since the maximal
contribution to the SZ effect at a frequency 217 GHz comes from the
plasma with temperature close to $k_{\mathrm{b}} T_{\mathrm{r}}=
160$ keV, we fix this temperature and derive the dependence of the
location of a ring on time
\begin{equation}
r=\frac{A}{T^{(\gamma-1)/3}_r} \left(\frac{t^{2 (2-\gamma)/5}
E^{(2\gamma+1)/15}}{\rho^{(2\gamma+1)/15}_1}\right)
\end{equation}
where
\begin{equation}
A=\left(\frac{8 (\gamma-1) m_{\mathrm{p}} \times
\beta^{(2\gamma+1)/(\gamma-1)} }{25 k_{\mathrm{b}}
(\gamma+1)^2}\right)^{(\gamma-1)/3}.
\end{equation}
If the adiabatic index equals $\gamma=5/3$ then the radius of a ring
is given by
\begin{equation}
r=\frac{A}{T^{2/9}_r} \left(\frac{t^2
E^{13/3}}{\rho^{13/3}_1}\right)^{1/15} \label{ringradius}
\end{equation}

Note that the ring radius depends weakly on time $r\propto t^{2/15}$
and represents a parameter which changes negligibly with time (in
contrast to the shock radius, see Eq. \ref{Rsedov}). The radius of a
ring can be used as one of the parameters for describing the Sedov
solution. If the ring radius is used instead of the Mach number then
the value of the released energy $E$ is derived from Eqs.
(\ref{Rsedov}), (\ref{ringradius}), depends strongly on the ring
radius,  and is given by
\begin{equation}
E\propto \frac{r^{9/2} \rho_1}{R^{3/2}}.
\end{equation}
Since the ring radius is $r(t)\propto t^{2/15}$ and the distance of
the shock from the origin is $R(t)\propto t^{2/5}$, the ratio of
$r^3(t)$ to $R(t)$ is constant in time and the value of the energy
is essentially determined by the ring radius of the SZ signal at a
frequency 217 GHz.

The analytical solution of hydrodynamic problems such as  the Sedov
solution is useful, but many problems have no such  solution and
numerical simulations are needed. Furthermore, the moderate values
observed for Mach numbers of shocks require inclusion of the effects
of gravity and finite initial gas pressure in the pre-shock regions.
We consider the SZ effect by relativistic electrons residing in AGN
cocoons during the active jet phase by using the results of
numerical simulations in Sect. 4.

\subsection{Synthetic observations of the SZ effect on AGN cocoons
in galaxy clusters}

High-resolution observations of the SZ effect are a promising tool
for observing AGN cocoons in galaxy clusters. Powerful AGN outbursts
were revealed in the Hydra A cluster (Nulsen et al. 2005b), in the
Hercules A cluster (Nulsen et al. 2005a), and in the MS0735.6+7421
cluster (McNamara et al. 2005). These clusters are located at the
redshifts of z=0.05, 0.15 and 0.22, respectively. A flat
$\Lambda$CDM cosmology, with $H_{0}=70$ km s$^{-1}$ Mpc$^{-1}$ and
$\Omega_{\mathrm{M}}=0.27$, gives a scale of 2.67 kpc arcsec$^{-1}$
for the redshift of z=0.15. Therefore, with the spatial resolution
of ALMA (2.8$''$ at 110 GHz), we will be able to resolve the
structures down to 10 kpc at z=0.15. In this section we simulate
observations of the ring-like structure on the SZ intensity map
studied in the previous section.

\begin{figure}[h]
\centering
\includegraphics[angle=90, width=8cm]{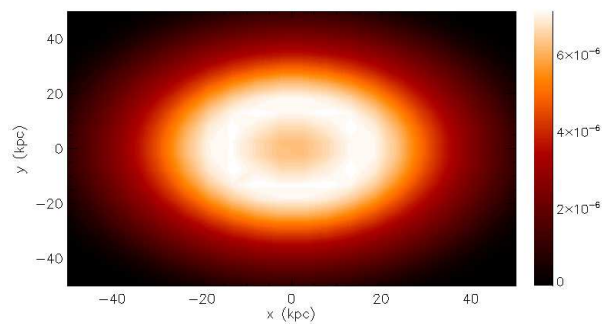}
\caption{Synthetic observation of an AGN outburst in a galaxy
cluster. The intensity map $|\Delta I/ I_0|$ of the SZ effect at a
frequency 217 GHz derived from the Sedov solution in the framework
of the Wright formalism and smoothed to the resolution of
FWHM$\simeq3.0''$.} \label{SZresol}
\end{figure}

An ALMA observation of the central region of the Perseus cluster is
simulated by Pfrommer et al. (2005) to reveal a presence of cavities
on the SZ intensity map. We convolve the SZ intensity map with a
Gaussian to obtain the resolution of the ALMA compact configuration
as it was done by Pfrommer et al. (2005). Since the angular extent
of a 100 kpc box (considered in Sect. 3) at the redshift z=0.15 is
close to the ALMA field of view ($\approx 30''$), for the sake of
simplicity we put a synthetic AGN cocoon in a galaxy cluster at the
redshift z=0.15. The resulting synthetic SZ intensity map of an AGN
cocoon at a frequency of 217 GHz in a galaxy cluster at z=0.15 is
shown in Fig. \ref{SZresol}. Figure \ref{SZresol} shows that the
ring-like structure is still present on the SZ intensity map after a
convolution with Gaussian of FWHM$\simeq3.0''$ and, therefore, the
ring radius measurement will be an interesting method provided by
the next generation telescopes (e.g. ALMA) to determine the energy
released during an outburst. The spatial resolution of $3''$ is
sufficient to resolve the ring-like structure on the SZ intensity
map at frequency of 217 GHz and, therefore, ALMA with the spatial
resolution of $1.3''$ at a frequency of 217 GHz in the most compact
configuration should reveal the ring-like structures on SZ intensity
maps. If we choose the same AGN cocoon in a cluster at z=0.05, the
spatial resolution of 10$''$ is sufficient to reveal the ring-like
structure on the SZ intensity map.

\makeatletter
\def\url@leostyle{\@ifundefined{selectfont}{\def\UrlFont{\sf}}{\def\UrlFont{\tiny\ttfamily}}}
\makeatother \urlstyle{leo}

We simulated SZ maps for ALMA with the spatial resolutions of $3''$
and $1.3''$ at frequencies of 90 GHz and 217 GHz, respectively. Our
simulated SZ intensity maps for ALMA are similar to those in Figs.
\ref{sz90} and \ref{sz217}, and available for download at the
webpage \url{http://www.oact.inaf.it/cosmoct/web\_group/
 research4/cocoon\_sz.html}

Pfrommer et al. (2005) examined whether the plasma bubbles are
detectable by ALMA and predicted 5 $\sigma$ detection of the bubble
of the Perseus cluster in an exposure time of 5 hours. Note that a
signal at a frequency of 217 GHz derived from the Sedov solution is
four times smaller than that analyzed by Pfrommer et al. (2005) at a
frequency of 144 GHz (see Figs. \ref{sz217} of this paper and 1 of
Pfrommer et al. 2005). The signal-to-noise ratio for a detection of
an AGN cocoon is proportional the square root of the exposure time
(e.g. Pfrommer et al. 2005; for a review, see a document
\footnote{http://www.cv.nrao.edu/naasc/ALMAsensitivity.ps}) and,
therefore, to detect the signal at a frequency of 217 GHz the ALMA
exposure time of 80 (i.e. $5\times4^2$) hours is required. The ALMA
Design Reference Science Plan (Version 2.2) contains two
projects\footnote{http://www.eso.org/sci/facilities/alma/science/drsp/}
in SZ studies (1.4.1 and 1.4.2) with integration time for each
program of 400 hours and new projects concern the SZ effect on AGN
cocoons should be promising. The Chandra X-ray satellite with a high
spatial resolution of 0.5$''$ has studied a non-relativistic
electron component in the ICM, while ALMA high-resolution
observations of AGN cocoons in galaxy clusters should permit us to
reveal a new (mildly relativistic) electron component.

\section{The SZ effect from a simulated cocoon}

Recent hydrodynamic simulations do not take into account
relativistic corrections to the SZ effect to derive SZ intensity
maps (e.g. Sijacki et al. 2008). Such relativistic corrections are
necessary for calculating the SZ effect on AGN cocoons produced by
electrons with high temperatures. So far, studies of the SZ effect
in the relativistic formalism were done only for an analytic toy
cocoon model (see Pfrommer et al. 2005).  In this section, we derive
the SZ intensity map by using hydrodynamical simulations and the
Wright relativistic formalism.

In previous papers (Antonuccio-Delogu \& Silk 2008; Tortora et al.
2009),  we used an Adaptive Mesh Refinement (AMR) code to follow the
evolution of the cocoon produced by the jet propagating in the
interstellar (ISM)/intergalactic medium (IGM). To perform the
simulation, we used FLASH v.2.5 (Fryxell et al. 2000), a parallel,
AMR code, which implements a second order, shock-capturing,
Piecewise-Parabolic-Method (PPM) solver. The jet is modelled as a
one-component fluid, characterized by a density which is a small
fraction of the initial density of the ISM. In this simulation, the
power injected by the jet is $10^{46}$ erg/s. We model the
environment, where the jet propagates, as a two-phase ISM,
comprising a hot, diffuse, low-density component having a
temperature and a central density at $10^7$ K and $1$ cm$^3$,
respectively, and a cold, clumped system of high-density clouds in
pressure equilibrium with the diffuse component. Such values of
temperature and pressure are typical of the ISM in the central parts
of an elliptical at high redshift. We assume that the diffuse gas is
embedded within a dark matter (DM) halo, the latter being described
by a Navarro-Frenk-White (NFW) density profile.

We choose a simulation box having a size $L_{\mathrm{box}} = 40
h_{-1}$ kpc, where $h$ is defined such that the Hubble constant
$H_0$ is 100 $h$ km s$^{-1}$ Mpc$^{-1}$. The spatial resolution
attained is a function of the maximum refinement level and of the
structure of the code. For a block-structured AMR code like FLASH,
where each block is composed by $n_x \times n_y$ cells, the maximum
resolution along each direction is given by $L_{\mathrm{box}}/(n_x
2^l)$, where $l$ is the maximum refinement level. In this
simulation, $n_x$ = $n_y$ = 8 and $l$ = 6, thus the minimum resolved
scale is 78.125 pc. Note that we are performing a 2D simulation, but
we do not impose any special symmetry.

\begin{figure}[h]
\centering
\includegraphics[angle=90, width=8cm]{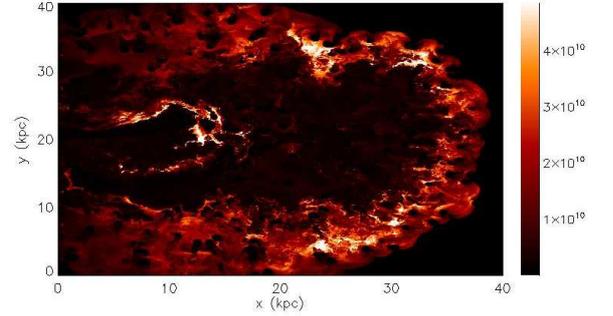}
\caption{The simulation map of the gas temperature (in K) in the AGN
cocoon} \label{Tsim}
\end{figure}

\begin{figure}[h]
\centering
\includegraphics[angle=90, width=9cm]{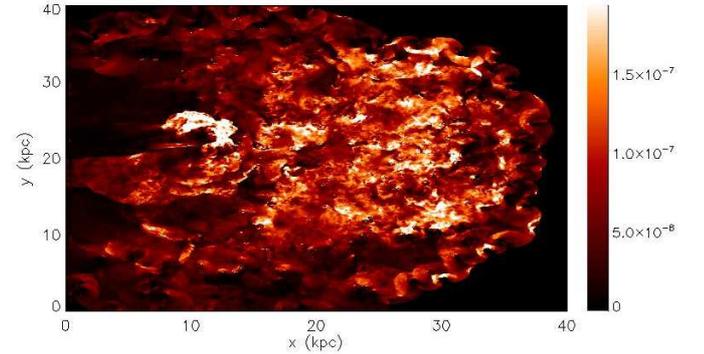}
\caption{The simulation map of the gas pressure (in erg/cm$^3$) in
the AGN cocoon} \label{Psim}
\end{figure}

Numerical simulations by Antonuccio-Delogu \& Silk (2008) and
Tortora et al. (2009) show that the temperatures can reach very high
values within the cocoons (T $\approx 10^9-10^{11}$ K), if the ISM
is dominated by a population of cold, star forming clouds embedded
into and in approximate pressure equilibrium with a hot, diffuse
phase. The pressure within the cocoon can reach high values because
the temperatures are on average very high. The SZ effect should be
significant since it determined by the pressure of the electron
population integrated along a line of sight. Our simulation
temperature and pressure maps are shown in Fig. \ref{Tsim} and
\ref{Psim}.

Figure \ref{Tsim} reveals a hot shell with a temperature of
$k_\mathrm{b} T_\mathrm{e}>2$ MeV around the internal region of the
simulation AGN cocoon. The temperatures of the internal region are
smaller than that of the hot shell and, therefore, absolute values
of the generalized spectral functions at frequencies of 90 and 217
GHz should be higher for the internal region in the corresponding
temperature range (see Figs. \ref{Fig90}, \ref{Fig217}, and
\ref{Tsim}).

We find that the pressure values are higher inside the internal
region in comparison with the thermal hot shell pressure. Therefore,
the SZ effect in the cocoon internal region should be more
significant than that produced by electrons located inside the hot
shell. The pressure inside the simulation cocoon is two orders of
magnitude higher than that found in the diffuse ISM. Such a
pressure-jump corresponds to the Mach number of a shock equaled
$M\approx\sqrt{4 P_2/5 P_1}\sim 10$ (Landau \& Lifshitz 1959). The
plasma inside the AGN cocoon is largely over-pressured relative to
the ambient ISM. The radio lobes of the nearby radio galaxy
Centaurus A is expanding into the ISM at the Mach number of 8.5
(Kraft et al. 2003) is representative of such over-pressured
plasmas.

\begin{figure}[h]
\centering
\includegraphics[angle=90, width=8cm]{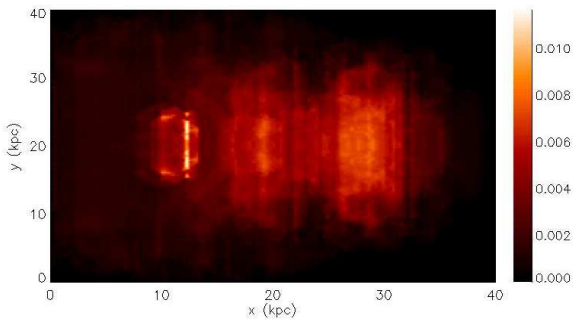}
\caption{The intensity map $|\Delta I/I_0|$ of the SZ effect at a
frequency 217 GHz derived from the numerical simulation in the
framework of the Wright formalism} \label{sz-sim}
\end{figure}

To produce the 3D pressure and temperature maps which are necessary
to derive the intensity map of the SZ effect we rotate the 2D
pressure and temperature maps along the jet axis. We calculated the
SZ effect using the values of the generalized spectral functions
found in  Sect. 2. Since the gas temperatures in a cocoon are high,
such a cocoon should be a source of the SZ effect at a frequency 217
GHz. The intensity map of the SZ effect at a frequency 217 GHz
derived from the simulation maps of the gas pressure and temperature
is plotted in Fig. \ref{sz-sim}.

A comparison of Figs. \ref{Tsim} and \ref{Psim} with Fig.
\ref{sz-sim} strongly suggests that the dominant contribution to the
SZ effect at a frequency 217 GHz originates from the cocoon internal
region. Therefore, measurements of the SZ effect at a frequency 217
GHz are a powerful tool for potentially revealing the dynamically
dominant component inside AGN cocoons (see Fig. \ref{Psim}). The
proposed method based on the generalized spectral function is a way
of maximizing the contribution from the gas with a temperature in
the range $10^9$ K $< T_{\mathrm{e}}< 10^{10}$ K (see Fig.
\ref{Fig217}). The study of the SZ effect in a plasma at such
temperatures is important since numerical simulations predict that
such a plasma represents the still invisible dynamically dominant
component inside AGN cocoons.

Our simulations are characterized by ISM density and temperature at
1 cm$^{-3}$ and $10^7$ K (ISM pressure at $\sim 10^{-9}$
erg/cm$^3$), respectively, typical of the ISM in the central parts
of an elliptical at high redshift ($z\approx1$). We found that the
plasma inside the AGN cocoon is largely over-pressured relative to
the ambient ISM. The derived intensity $\Delta I$ of the SZ effect
in this case is three orders of magnitude higher than that of
cocoons in galaxy clusters. Though the linear scales of cocoons are
close in both cases, the angular diameter of the cocoon within the
distant elliptical galaxy equaled to
$\theta_{\mathrm{z}=1}=L_{\mathrm{box}}/D_\mathrm{A}\sim
2.5^{\prime\prime}$ is much smaller than that of the cocoon with the
same linear size inside the Perseus cluster
$\theta_{\mathrm{z}=0.0179}=1.8^{\prime}$, the angular distance is
denoted by $D_\mathrm{A}$. Then the SZ fluxes from such cocoons
which are proportional to $\Delta I/ D^2_\mathrm{A}$ should be of
the same order (see Figs. \ref{sz217} and \ref{sz-sim}).

The main difference in the intensity of the signal is then a result
of the higher internal pressure in the simulated cocoon. The
pressure of high temperature gas in the cavity center, in the
framework of the analytical model in Sect. 3, can not be much higher
than the ambient pressure of a galaxy cluster $\sim3\times10^{-11}$
erg/cm$^3$ because of the moderate Mach number of the shock of $M=2$
and a decrease of the pressure from the shock front to the cavity
center (e.g., Sedov 1959).

\subsection{An analysis of the hot gas by means of the combined
generalized function}

As was shown in Sect. 2.2 using the combined generalized spectral
functions provides us with an alternative to a measurement of the SZ
effect at a frequency 217 GHz for studying a population of electrons
with energies higher than 100 keV. In this section we derive the
intensity map of the SZ effect by means of the combined function
$C(2.26, 6.51, T_{\mathrm{e}})$ and consider how the contribution
from the kinematic Sunyaev-Zel'dovich (SZ) effect  can be eliminated
(for review of the kinematical SZ effect, see Birkinshaw 1999).

\begin{figure}[h]
\centering
\includegraphics[angle=90, width=8cm]{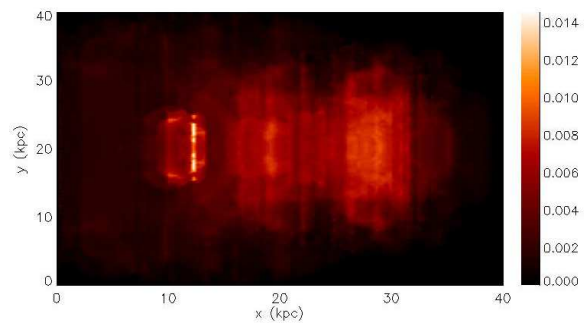}
\caption{The intensity map $|\Delta I/ I_0|$ of the SZ effect
derived from our numerical simulation by using the combined
generalized spectral function $C(2.26, 6.51, T)$} \label{sz-sim2}
\end{figure}

The generalized spectral function at a frequency 217 GHz belongs to
the family of spectral functions defined by Eq. (\ref{C}). Both
$G(x, T_{\mathrm{e}})$ and $C(2.26, 6.51, T_{\mathrm{e}})$ functions
have absolute maximal values at temperatures above 100 keV and,
therefore, these functions are interesting for analyzing the
presence of high energy electrons which are expected to be inside
AGN cocoons. The intensity map $\Delta I/ I_0$ of the SZ effect
derived from our numerical simulation by using the combined
generalized spectral function $C(2.26, 6.51, T)$ is plotted in Fig.
\ref{sz-sim2}.

The similarity in shape between the SZ intensity maps in Figs.
\ref{sz-sim} and \ref{sz-sim2} shows that a population of high
energy electrons can be constrained by means of different and
equivalent methods based on measurements of the SZ effect.

A source of bias in the observations of the SZ effect by
energetic electrons could be provided by a possibly relevant
kinematic SZ effect. The kinematic SZ effect arises from the bulk
motion of the medium relative to the CMB rest frame. The change of
the CMB intensity due to the kinematic SZ effect is given by
\begin{equation}
\Delta I_{\mathrm{kin}}(x)=-\beta \tau h(x), \label{Ikin}
\end{equation}
where $\beta$ is the line-of-sight velocity of bulk motion in units
of the speed of light. The spectral function $h(x)$ is given by
\begin{equation}
h(x)=\frac{x^4 \exp(x)}{[\exp(x)-1]^2}.
\end{equation}

The absolute maximum of the kinematic SZ effect lies at a frequency
217 GHz which coincides with the crossover frequency of the
non-relativistic thermal SZ effect. Pfrommer et al. (2005) concluded
that it is impossible to remove the degeneracy of the cocoon optical
depth and the kinematic SZ effect using only a single frequency SZ
observation. Colafrancesco et al. (2009) found that the value of the
slope $S=(\Delta I(x_{\mathrm{a}})-\Delta
I(x_\mathrm{b}))/(x_{\mathrm{a}}-x_{\mathrm{b}})$ of the SZ effect
does not depend on the kinematic SZ effect spectrum in the frequency
range around the crossover frequency of the thermal SZ effect, i.e.
in the frequency range $x_\mathrm{a}, x_\mathrm{b} \in (3.5, 4.5)$.
We now show that there is another way to eliminate the kinematic SZ
effect based on the combined generalized spectral functions.

Let us define the intensity function $U$ as
\begin{eqnarray}
&&U=\Delta I(x_{\mathrm{min}})-\Delta
I(x_{\mathrm{max}})\times\frac{g(x_{\mathrm{min}})}{g(x_{\mathrm{max}})}-\label{U}\\
&&\Delta I(x_0)\times\left(\frac{h(x_{\mathrm{min}})}{h(x_0)}
-\frac{h(x_{\mathrm{max}}) }{h(x_0)}\times\frac{g(x_{\mathrm{min}})}
{g(x_{\mathrm{max}})}\right), \nonumber
\end{eqnarray}
where $\Delta I(x)$ is the total intensity of the thermal and
kinematic SZ effects, $x_{\mathrm{min}}=2.26$ ($\nu$ = 128 GHz),
$x_0=3.83$ ($\nu$=217 GHz), and $x_{\mathrm{max}}=6.51$ ($\nu=369$
GHz) correspond to the frequencies of the minimum, the crossover and
the maximum of the non-relativistic thermal SZ effect, respectively.

\begin{figure}[h]
\centering
\includegraphics[angle=0, width=7cm]{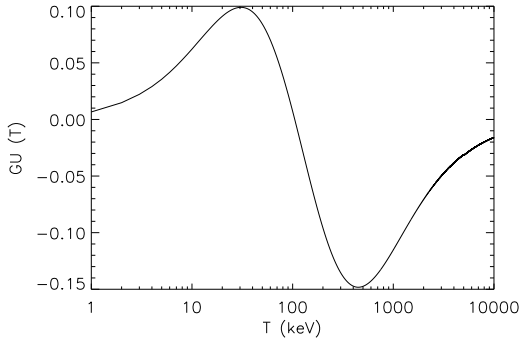}
\caption{Dependence of the function $GU(T_{\mathrm{e}})$ at on the
plasma temperature.} \label{GU}
\end{figure}

Since $\Delta I(x_{\mathrm{min}})-\Delta I(x_{\mathrm{max}})\times
g(x_{\mathrm{min}})/g(x_{\mathrm{max}})$ does not depend on the
non-relativistic thermal SZ effect (see Sect. 2.2) and $g(x_0)$
equals zero, there is no contribution of the non-relativistic
thermal SZ effect to the intensity function $U$. Note that $\Delta
I(x_{\mathrm{min}})-\Delta I(x_0)\times h(x_{\mathrm{min}})/h(x_0)$
and $\Delta I(x_{\mathrm{max}})-\Delta I(x_0)\times
h(x_{\mathrm{max}})/h(x_0)$ have no dependence on the kinematic SZ
effect, and therefore the study of the intensity function $U$ is a
way of constraining a population of high energy electrons without
biases provided by a possible relevant kinematic SZ effect. The
value of the last bracket on the right-hand side of Eq. (\ref{U}) is
very close to 1 (it equals 1.029) and, therefore, the spectral
function $GU(T)$ which corresponds to the intensity function $U$
(see Eq. \ref{form}) is approximately equal to
\begin{equation}
GU (T_{\mathrm{e}})\approx C(2.26, 6.51, T_{\mathrm{e}})-G(3.83,
T_{\mathrm{e}})
\end{equation}
where the spectral functions $C(2.26, 6.51, T_{\mathrm{e}})$ and
$G(3.83, T_{\mathrm{e}})$ correspond to the SZ intensities of
$\Delta I(x_{\mathrm{min}})-\Delta I(x_{\mathrm{max}})\times
g(x_{\mathrm{min}})/g(x_{\mathrm{max}})$ and $\Delta I (x_0)$,
respectively. The spectral function $GU(T)$ is shown in Fig.
\ref{GU}.

The intensity map $\Delta I/I_0$ of the SZ effect derived from our
numerical simulation by using the function $GU(T)$ corresponds to a
residual between the intensity maps of Figs. \ref{sz-sim2} and
\ref{sz-sim}. The function $GU(T)$ has the minimum at temperatures
higher than 300 keV and is not affected by the kinematic SZ effect,
and therefore differencing these SZ intensity maps (see Fig.
\ref{sz-sim} and \ref{sz-sim2}) is an important tool for analyzing a
high energy electron population.

In the next section, we consider the scattered CMB photon spectrum at
higher frequencies. This study is important since photon energy
gains due to the inverse Compton effect on high energy electrons are
substantial.

\section{Flatness of the high-frequency tail of the scattered photon spectrum}

Electrons with high energies produce a substantial energy gain of
photons, which scattered to the high-frequency tail. For an optical
depth $\tau\ll1$ and a sufficiently large Comptonization parameter
$y$, a sequence of declining peaks in the high-frequency tail
($x\gg10$) of the scattered photon energy flux spectrum is present
(e.g., see Loeb et al. 1991). These peaks correspond successively to
singly, doubly (etc) scattered photons (see Figs. 4a-4d in Loeb et
al. 1991). Next,  we study the high-frequency tail of the scattered
CMB spectrum by high energy electrons at frequencies below the first
peak frequency.

At high frequencies (i.e. $x\gg10$) the term $x^3/(\exp(x)-1)$ of
 Eq. (\ref{G}) decreases strongly and, therefore, the Eq.
(\ref{G}) can be written as
\begin{equation}
G(x,T_{\mathrm{e}})\approx \int^{\infty}_{-\infty} \frac{P_1 (s,
T_{\mathrm{e}})}{\Theta(T_{\mathrm{e}})} \frac{x^3 \exp(-3
s)}{\exp(x \exp(-s))-1} ds.
\end{equation}
The treatment is considerably simplified if $\exp(x \exp(-s))\gg1$
and in the case the generalized spectral function $G(x,
T_{\mathrm{e}})$ is given by
\begin{equation}
G(x,T_{\mathrm{e}})\approx \int^{\infty}_{-\infty} \frac{P_1 (s,
T_{\mathrm{e}})}{\Theta(T_{\mathrm{e}})} x^3 \exp\left(-3 s-x
\exp(-s)\right) ds.\label{Ghf}
\end{equation}

The sub-exponential function $f(s)=-3 s-x \exp(-s)$ has a maximum at
the frequency shift $s_{\mathrm{max}}=\ln(x/3)$. Since $\exp(x
\exp(-s_{\mathrm{max}}))\gg1$ the approximate expression for the
generalized spectral function is valid and we calculate the integral
in Eq. (\ref{Ghf}) by the Laplace's method. The approximate value of
the sub-exponential function in a neighborhood of the point
$s=s_{\mathrm{max}}$ is then
\begin{equation}
f(s)=-3s_{\mathrm{max}} -x e^{-s_{\mathrm{max}}}-\frac{x}{2}
e^{-s_{\mathrm{max}}}\times (s-s_{\mathrm{max}})^2
\end{equation}
and the spectral function approximately equals
\begin{equation}
G(x, T_{\mathrm{e}})\approx \sqrt{\frac{2\pi}{3}}
\left(\frac{3}{e}\right)^3 \frac{P_1\left(\ln\frac{x}{3},
T_{\mathrm{e}}\right)}{\Theta(T_{\mathrm{e}})}.
\end{equation}

\begin{figure}[h]
\centering
\includegraphics[angle=0, width=7cm]{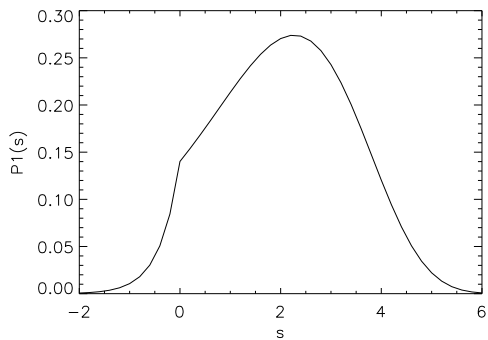}
\caption{Dependence of the distribution of frequency shifts $P_1 (s,
k_{\mathrm{b}} T_{\mathrm{e}})$ at the temperature of 500 keV.}
\label{P1}
\end{figure}

As was shown by Kino et al. (2007) and Antonuccio-Delogu \& Silk
(2008), high gas temperatures ($k_{\mathrm{b}} T_{\mathrm{e}}\sim1$
MeV) are expected in AGN cocoons. For example, we choose the
temperature equaled to 500 keV to study the high-frequency tail of
the generalized spectral function.
For high temperatures the function $P_1 (s, T_{\mathrm{e}})$ is wide
and it is centered at high values of the frequency shift $s$. The
distribution of frequency shifts for single scattering $P_1 (s,
T_{\mathrm{e}})$ at the temperature of $k_{\mathrm{b}}
T_{\mathrm{e}}=500$ keV is shown in Fig. \ref{P1}.

Figure \ref{P1} shows that the values of the distribution of
frequency shift lie in the narrow range (0.25, 0.28) if the
frequency shift $s$ is in the range (1.5, 3.0). The frequency shift
$s_{\mathrm{max}}$ lies in this range when the dimensionless
frequency is of $15<x<90$ and, therefore, the approximate value of
the generalized spectral function in this frequency range is
\begin{equation}
G(x, 500 \mathrm{keV})\approx 0.25\sqrt{\frac{2\pi}{3}}
\left(\frac{3}{e}\right)^3 \frac{1}{\Theta(500 \mathrm{keV})}.
\end{equation}

\begin{figure}[h]
\centering
\includegraphics[angle=0, width=7cm]{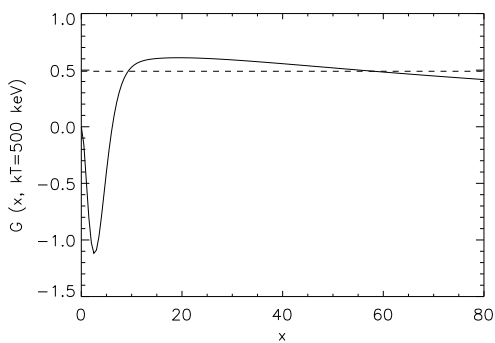}
\caption{Dependence of the spectral function $G\left(x,
k_{\mathrm{b}} T_{\mathrm{e}}\right.=500$ keV) on the dimensionless
frequency $x$ is shown by the solid line. The approximate value of
$G\left(x, k_{\mathrm{b}} T_{\mathrm{e}}\right.=500$ keV) is shown
by the dashed line.} \label{500}
\end{figure}

The quantitative dependence of the spectral function $G(x,
k_{\mathrm{b}} T_{\mathrm{e}}=500$ keV) on the dimensionless
frequency $x$ is illustrated in Fig. \ref{500}.

We conclude that the generalized spectral function should be flat in
the broad frequency range when the temperature values are
sufficiently high. Since the non-relativistic spectral function
$g(x)$ is a rapidly decreasing function at high frequencies $(x>10)$
in contrast with the spectral function $G(x, T_{\mathrm{e}})$ at
high electron temperatures, a measurement of the SZ effect at high
frequencies provides an interesting test of the presence of high
energy electrons.

\section{Discussion and conclusions}

The direct detection of the SZ effect can in principle provide a
unique diagnostic tool to study the physical conditions of the ICM
at high redshift. To reach this end, an accurate treatment of the
spectral properties of the SZ signal is very important, and this was
our main aim in this work. Previous models were based on simplifying
assumptions, like the assumption of pressure equilibrium and density
homogeneity (Colafrancesco, 2005; Pfrommer et al., 2005). Here we
have instead considered the SZ signal arising from a inhomogeneous
exactly solvable configuration, i.e. a spherically symmetric
Sedov-expanding region, and even more realistic models obtained from
2D fluid-dynamical simulations of jet/cocoon system propagating into
the ISM$/$IGM.\\
\noindent In the standard model of a typical strong double radio
sources, a couple of twin jets advances into the IGM and engenders a
hot, low-density ``cocoon'' which is filled only of matter
originating from the jet and of part of the shocked IGM (Scheuer
1974; Blandford \& Rees 1974). Numerical simulations by
Antonuccio-Delogu \& Silk (2008), Tortora et al. (2009) and
theoretical works by Kino et al. (2007) suggest that the plasma
temperatures within the cocoons are very high
($T\approx10^9-10^{11}$ K) and, therefore, these plasmas are
invisible in the soft X-ray band and generate cavities on the X-ray
maps (e.g. McNamara et al. 2005).

The energy density of the cocoon is higher than that of the IGM: it
is a local hot region that induces a significant thermal SZ effect
(Yamada et al. 1999). Since the temperature within the cocoons is
expected to be very high, we have adopted a relativistically
corrected formalism to calculate the SZ effect. Using the
relativistically corrected Wright formalism, Colafrancesco (2005)
and Pfrommer et al. (2005) have shown that the hot plasma within the
X-ray cavities should be observable by detecting the SZ signal at a
frequency $\nu=$217 GHz, since the contribution of the
non-relativistic IGM to the SZ effect at this frequency is
negligible.

However, as stated above, the analytical models for the cocoons considered by
Colafrancesco (2005) and Pfrommer et al. (2005) are based on the assumptions of pressure equilibrium and cocoon
homogeneity. On the other hand, recent numerical simulations of the
SZ effect do not include the relativistically corrected SZ treatment
(e.g. Sijacki et al. 2008). To produce more realistic SZ maps we
have considered the analytical Sedov model and numerical 2-D simulations
and calculate the SZ effect by using the Wright formalism.

We have studied the CMB distortion due to the SZ effect as a function of
the gas temperature. We have shown that the CMB spectral distortion can be expressed in
terms of the generalized spectral functions $G(x, T_{\mathrm{e}})$
which are functions of the gas temperature. This spectral function at the frequency 217 GHz has a peak at
a temperature $k_{\mathrm{b}} T_{\mathrm{e}}=160$ keV and,
therefore, analysis of the SZ effect at this frequency should be a
powerful way of checking for the existence of electron populations at high
energies.\\
\noindent We have also proposed a new method for excluding the contribution from the low
energy, non-relativistic electrons to the SZ effect by means of
observations at two frequencies. The derived
similarity of the generalized spectral function $G(3.83,
T_{\mathrm{e}})$ and of the combined generalized function $C(3.83,
6.51, T_{\mathrm{e}})$ provides us with an alternative and
equivalent method for studying a population of electrons with
energies higher than 100 keV. We have shown how one may correct
for a possible contribution from the kinematic SZ effect using the
difference between the generalized spectral function $G(3.83,
T_{\mathrm{e}})$ and the combined generalized function $C(3.83,
6.51, T_{\mathrm{e}})$.

In Sect. 3 we have studied in detail the SZ effect for a specific
astrophysical important case - the Sedov self-similar solution. The
Sedov solution was used by McNamara et al. (2005) to described the
observed X-ray cavities. We have then considered the AGN cocoon in a
galaxy cluster, and calculated the SZ effect at a frequency 217 GHz
produced by the hot gas inside an X-ray cavity. We have used the
radius of the ring structure on the SZ intensity map for deriving
the value of the released energy. In Sect. 3.2 we have also showed
that the ALMA telescope is needed to reveal mildly relativistic
electron populations in AGN cocoons.

We have then proceeded by considering the SZ effect from a simulated cocoon in
Sect. 4. The simulated cocoon is largely over-pressured relative to
the ambient ISM. We have derived the SZ intensity map by using
hydrodynamical simulations and the Wright relativistic formalism. We
showed that the SZ fluxes from cocoons in the central part of a
distant elliptical and a nearby galaxy cluster are of the same
order.

We have showed that the CMB spectral distortion is flat in a broad range of high
frequencies if gas temperatures are high, and we have estimated the
approximate value of the generalized spectral function at
temperature 500 keV in the frequency range $15<x<90$.

Our study demonstrates that the SZ effect, originating from
populations of high energy electrons residing in AGN cocoons,
calculated using the SZ intensity maps in the framework of
relativistically corrected formalism, can give a detectable signal.
We have proposed methods for maximizing the contribution of high
energy electrons, which provides promising tests for finding the
hitherto invisible hot component of AGN cocoons. The detections and
measurements of these populations are within the reach of ALMA, and
could provide a fundamental tool to characterize the evolution of
AGN activity at high redshifts.

\appendix

\section{Fluid dynamical scaling of cocoon}
In the Appendix A we show how one can produce intensity maps of the
SZ effect by means of a dynamical scaling of cocoon. We will use a
slightly modified version of the scaling fluid dynamical equations
by Tang \& Wang (2009), which applies to the case of an expanding
cocoon continuously powered by a jet. The set of equations we solve
is very similar to Eqs. (1)-(4) of Tang \& Wang (2009), with the
only difference that the energy equation now contains a source term
describing the injection of energy density:
\begin{equation}
\frac{\partial\rho e}{\partial t} + \nabla\cdot\left[\left(\rho e +
P\right)\mathbf{v}\right] = J \delta(\vec{r}-\vec{r_{\mathrm{b}}}) -
\hat{L} \label{eq:app1}
\end{equation}
where: $J$ is the power injected by the jet into the cocoon,
$\delta(\vec{r}-\vec{r_{\mathrm{b}}})$ denotes a 3-dimensional
Dirac's delta function with $\vec{r_{\mathrm{b}}}$ denoting the
(bounded) jet's injection region, and $\hat{L} =
n_{\mathrm{e}}n_{\mathrm{H}}\Lambda(T)$ is the cooling term. Tang
and Wang (2009) assume a scaling of the form: $q_{2} =
q_{1}Q^{i_{\mathrm{q}}}$, where: $q = (L,M,t,T,v,e,\rho,p,E,J)$ is
either a fundamental or derived physical quantity: length $L$, mass
$M$, time $t$, temperature $T$, velocity $v$, specific energy $e$,
density $\rho$, pressure $p$,  total energy $E$, and jet's power
$J$) and $Q$ is a positive constant. Under the hypothesis that the
cooling function $\Lambda \simeq T^{1/2}$, which holds for $T \geq
5\times10^{6}\, \rm{K}\,$ in a fully ionized plasma, the scaling
quantities are found to be connected by the following relations:
\begin{equation}
i_{\mathrm{v}} = i_{\mathrm{L}} - i_{\mathrm{t}}
\end{equation}
\begin{equation}
i_{\mathrm{e}} = 2i_{\mathrm{L}} - 2i_{\mathrm{t}}
\end{equation}
\begin{equation}
i_{\rho} = i_{\mathrm{M}} - 3i_{\mathrm{L}} \label{irho}
\end{equation}
\begin{equation}
i_{\mathrm{p}} = i_{\mathrm{M}} - i_{\mathrm{L}} - 2i_{\mathrm{t}}
\end{equation}
\begin{equation}
i_{\mathrm{E}} = i_{\mathrm{M}} + 2i_{\mathrm{L}} - 2i_{\mathrm{t}}
\end{equation}
\begin{equation}
i_{\mathrm{T}} = 2i_{\mathrm{L}} - 2i_{\mathrm{t}} \equiv
i_{\mathrm{e}}
\end{equation}
\begin{equation}
i_{\mathrm{M}} = -3i_{\mathrm{t}} + 5i_{\mathrm{L}} -
0.5i_{\mathrm{T}} \label{eq:cool}
\end{equation}
\begin{equation}
i_{\mathrm{J}} = i_{\mathrm{M}} + 2i_{\mathrm{L}} - 3i_{\mathrm{t}}
\label{eq:pjet}\label{iw}
\end{equation}
Eq.~\ref{eq:cool} was derived from the scaling of the cooling term,
and the final equation (eq.~\ref{eq:pjet}) derives assuming the
following scaling of jet's power:
$J \propto J_{0} Q^{i_{\mathrm{J}}}$.\\
Note that Eq.~\ref{eq:cool} is different from Eq. (15) of Tang \&
Wang (2009). The latter reads: $i_{M} = -3i_{t} + 5i_{L} +
0.5i_{T}$. Our derivation is obtained starting from the scaling of
the right-hand side terms: $\partial\left(\rho e\right)/\partial t
\sim i_{\rho} + i_{e} - i_{t} = i_{M} -3i_{L} + 2(i_{L} - i_{t}) -
i_{t} = i_{M} - i_{L} - 3i_{t}$, where the symbol ``$\sim$'' denotes
the power law scaling index. The number density scales as
$n\sim\rho$ and, therefore, the cooling term scales as $L \sim
2i_{M}-6i_{L} + 0.5i_{T}$. Thus, by equating the two terms, we have:
$i_{M} - i_{L} - 3i_{t} =2i_{M}-6i_{L} + 0.5i_{T}$, which gives
Eq.~\ref{eq:cool}.
\\
As described by Tang \& Wang (2009), this linear system is
completely specified after the choice of 2 of the coefficients. We
choose to fix the scaling coefficients for length and time, $i_{L}$
and $i_{t}$, and find
\begin{eqnarray}
\nonumber&&i_{\mathrm{v}} = i_{\mathrm{L}} - i_{\mathrm{t}}\\
\nonumber&&i_{\mathrm{e}} = i_{\mathrm{T}} = 2 i_{\mathrm{L}} - 2 i_{\mathrm{t}}\\
&&\nonumber i_{\rho} = i_L - 2 i_t\\
&&i_{\mathrm{p}} = 3 i_{\mathrm{L}} - 4 i_{\mathrm{t}} \label{sys}\\
&&\nonumber i_{\mathrm{E}} = 6 i_{\mathrm{L}} - 4 i_{\mathrm{t}}\\
&&\nonumber i_{\mathrm{M}} = 4 i_{\mathrm{L}} - 2 i_{\mathrm{t}}\\
&&\nonumber i_{\mathrm{J}} = 6 i_{\mathrm{L}} - 5 i_{\mathrm{t}}.
\end{eqnarray}

Thus, a given solution (also numerical) can be rescaled by means of
Eqs. (\ref{sys}) to a different jet's power using scalings
coefficients.

\begin{figure}[h]
\centering
\includegraphics[angle=90, width=8cm]{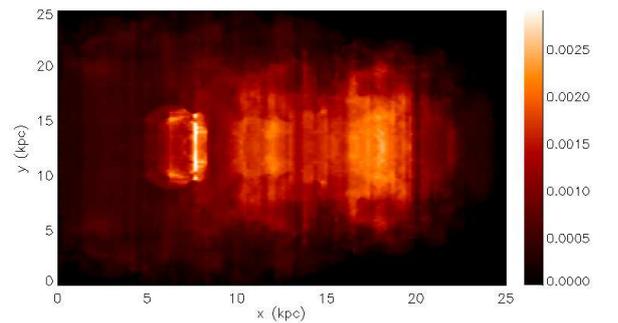}
\caption{The intensity map $|\Delta I/ I_0|$ of the SZ effect at a
frequency 217 GHz derived from the dynamical scaling.}
\label{sz-scaling}
\end{figure}

To produce an intensity map of the SZ effect by a cocoon surrounding
a less powerful jet using the dynamical scaling of cocoon we choose
scaling parameters equaled: $Q=10$, $i_L=-0.2$, and $i_t=-0.05$. The
scaled power injected by the jet is $10^{45}$ erg/s and the scaled
size of the simulation box is 25$h_{-1}$ kpc. The intensity map of
the SZ effect at a frequency 217 GHz derived from the scaling of the
simulation maps of the gas pressure and temperature is plotted in
Fig. \ref{sz-scaling}.

Thus a new intensity map of the SZ effect can be found for cocoons
evolving in different ISM by means of one particular simulation and
of the dynamical scaling. Therefore, the dynamical scaling is a
promising method to produce the intensity maps of the SZ effect by
hot gas residing in AGN cocoons in different ISM.

\section{Measurements at 217 GHz}

In Appendix B we show how the broad detector spectral response
impacts on the possibility of an analysis of high temperature
plasmas.

We convolve the spectral function $g(x)$ over a Gaussian detector
response centered at 217 GHz as it was done by Shimon \& Rephaeli
(2003). The signal at a frequency of 217 GHz from a low-temperature
plasma which arises due to the convolution over a spectral response
is shown in Fig. B.1 as a function of the FWHM of the Gaussian
detector response.

\begin{figure}[h]
\centering
\includegraphics[angle=0, width=8cm]{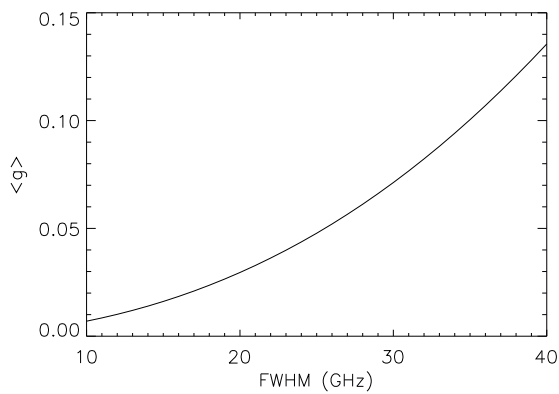}
\caption{The contribution of the SZ signal at a frequency of 217 GHz
due to the convolution over a Gaussian spectral response as a
function of the FWHM}
\end{figure}

The contribution of the SZ signal at a frequency of 217 GHz due to
the convolution over a spectral response with the FWHM $<30$ GHz is
far smaller than the SZ signal due to the presence of high
temperature plasmas (see Fig. \ref{Fig217}) and, therefore, the SZ
effect at frequency of 217 GHz is a promising tool for analyzing the
hot electron component in an AGN cocoon if the FWHM of a spectral
response is smaller than 30 GHz.

\begin{acknowledgements}
We are grateful to Kwang-il Seon for valuable discussions and thank
the referee for very useful comments.
\end{acknowledgements}


\begin{thebibliography}{99}
\bibitem{Antonuccio-Delogu 2008}
Antonuccio-Delogu, V., Silk, J. 2008, MNRAS, 389, 1750
\bibitem{Begelman 1989}
Begelman, M. C., Cioffi, D. F. 1989, APJ, 345, L21
\bibitem{Birkinshaw 1999}
Birkinshaw, M. 1999, Phys. Rep., 310, 97
\bibitem{Blandford 1974}
Blandford, R. D., Rees, M. J. 1974, MNRAS, 169, 395
\bibitem{Cavaliere 2006}
Cavaliere, A., Lapi, A. 2006, ApJ, 647L, 5
\bibitem{Chatterjee 2007}
Chatterjee, S., Kosowsky, A. 2007, ApJ, 661, L113
\bibitem{Colafrancesco 2005}
Colafrancesco, S. 2005, A\&A, 435, L9
\bibitem{Colafrancesco 2009}
Colafrancesco, S., Prokhorov, D. A., Dogiel, V. A. 2009, A\&A, 494,
1
\bibitem{Ensslin 2000}
En$\ss$lin, T. A., Kaiser, C. R. 2000, A\&A, 360, 417
\bibitem{Fabbri 1981}
Fabbri, R. 1981, Ap\&SS, 77, 529
\bibitem{Forman 2005}
Forman, W., Nulsen, P., Heinz, S. et al. 2005, ApJ, 635, 894
\bibitem{Fryxell 2000}
Fryxell B. et al., 2000, ApJS, 131, 273
\bibitem{Ito 2008}
Ito, H., Kino, M., Kawakatu, N. et al. 2008, ApJ, 685, 828
\bibitem{Kino 2009}
Kino, M., Kawakatu, N., Itoh, H. 2007, MNRAS, 376, 1630
\bibitem{Kompaneets 1957}
Kompaneets, A. C. 1957, Soviet Phys. - JETP, 4, 730
\bibitem{Kraft 2003}
Kraft, R. P., V\'{a}zquez, S. E., Forman, W. R. et al. 2003, ApJ,
592, 129
\bibitem{Landau 1959}
Landau, L.D., Lifshitz, E.M. 1959, Fluid Mechanics (Addison- Wesley
Reading)
\bibitem{Loeb 1991}
Loeb, A., McKee, C. F. and Lahav, O. 1991, ApJ, 374, 44
\bibitem{Markevitch 2007}
Markevitch, M., Viklinin, A. 2007, Phys. Rep., 443, 1
\bibitem{McNamara 2000}
McNamara, B. R., Wise, M., Nulsen, P. E. J., et al. 2000, ApJ, 534,
L135
\bibitem{McNamara 2005}
McNamara, B. R., Nulsen, P. E. J., Wise, M. W. et al. 2005, Nature,
433, 45
\bibitem{Nulsen 2005a}
Nulsen, P. E. J., Hambrick, D. C., McNamara, B. R. et al. 2005, ApJ,
625, L9
\bibitem{Nulsen 2005b}
Nulsen, P. E. J., McNamara, B. R., Wise, M. W. and David, L. P.
2005, ApJ, 628, 629
\bibitem{Ostriker 1988}
Ostriker, J. P., McKee, C. F. 1988, Rev. Mod. Phys., 60, 1
\bibitem{Platania 2002}
Platania, P., Burigana, G., De Zotti, G. et al. MNRAS, 337, 242
\bibitem{Pfrommer 2005}
Pfrommer, C., En$\ss$lin, T. A., Sarazin, C. L. 2005, A\&A, 430, 799
\bibitem{Prokhorov 200}
Prokhorov, D. A. 2010, A\&A, 509, 29
\bibitem{Rephaeli 1995}
Rephaeli, Y. 1995, ApJ, 445, 33
\bibitem{Rybicki 1979}
Rybicki, G. B., Lightman, A. P. 1979, Radiative Processes in
Astrophysics, New-York
\bibitem{Sedov 1959}
Sedov, L. 1959, Similarity and Dimensional Methods in Mechanics.
Academic, New York
\bibitem{Scheuer 1974}
Scheuer, P. A. G. 1974, MNRAS, 166, 513
\bibitem{Shimon 2003}
Shimon, M., Rephaeli, Y. 2003, New Astronomy, 9, 69
\bibitem{Sijacki 2008}
Sijacki, D., Pfrommer, C., Springel, V., En$\ss$lin, T. 2008, MNRAS,
387, 1403
\bibitem{Sunyaev 1980}
Sunyaev, R. A., Zel'dovich, Ya. B. 1980, ARA\&A, 18, 537
\bibitem{Tang 2005}
Tang, S., Wang, Q. D. 2005, ApJ, 628, 205
\bibitem{Tang 2009}
Tang, S., Wang, Q. D. 2009, MNRAS, 397, 2106
\bibitem{Tortora et al.
2009} Tortora, C., Antonuccio-Delogu, V., Kaviraj, S. et al. 2009,
MNRAS, 396, 61
\bibitem{Wright 1979}
Wright, E. L. 1979, ApJ, 232, 348
\bibitem{Yamada 1999}
Yamada, M., Sugiyama, N., Silk, J. 1999, ApJ, 522, 66
\bibitem{Zel'dovich 1969}
Zel'dovich, Ya. B., Sunyaev, R. A. 1969, Astrophys. Space Sci., 4,
301
\end{thebibliography}
\end{document}